# Crystal design of altermagnetism

Zhiyuan Zhou[1], Xingkai Cheng[2], Mengli Hu[2], Junwei Liu[2], Feng Pan[1], and Cheng Song[1,*]

[1]Key Laboratory of Advanced Materials (MOE), School of Materials Science and Engineering, Tsinghua University, Beijing 100084, China.

[2]Department of Physics, The Hong Kong University of Science and Technology, Hong Kong 999077, China.

*Corresponding author. Email: songcheng@mail.tsinghua.edu.cn

**Symmetry plays a fundamental role in condensed matter[1–8]. The unique entanglement between magnetic sublattices and alternating crystal environment in altermagnets[7,9–17] provides a unique opportunity for designing magnetic space symmetry. There have been extensive experimental efforts concentrated on tuning the Néel vector[7,13–17] to reconstruct altermagnetic symmetry. However, it remains challenging to modulate the altermagnetic symmetry through the crystal aspect. Here, the crystal design of altermagnetism is successfully realized, by breaking glide mirrors and magnetic mirrors of the (0001) crystallographic plane in CrSb films via crystal distortion. We establish a locking relationship between altermagnetic symmetry and the emergent Dzyaloshinskii-Moriya (DM) vectors in different CrSb films, realizing unprecedentedly room-temperature spontaneous anomalous Hall effect in an altermagnetic metal. The concept of exchange-coupling torques is broadened to include both antiferromagnetic exchange-coupling torque[18–20] and DM torque. Their relationship is designable, determining electrical manipulation modes, e.g., field-assisted switching for CrSb($1\bar{1}00$)/Pt and field-free switching for W/CrSb($11\bar{2}0$). Particularly, the**

**unprecedentedly field-free 100-percent switching of Néel vectors is realized by making these two torques parallel or antiparallel, dependent on Néel vector orientation. Besides unravelling the rich mechanisms for electrical manipulation of altermagnetism rooted in broadened concept of exchange-coupling torques, we list other material candidates and propose that crystal design of altermagnetism would bring rich designability to magnonics, topology, etc.**

The design and manipulation of symmetry is at the center of condensed matter.[1–8,23–26] As an unconventional antiferromagnetic phase combining the advantages of both ferromagnets and conventional antiferromagnets, altermagnets[7,9–17,27–33] is capable of generating time-reversal-symmetry-breaking (TRS-breaking) responses, while keeping the advantages of vanishing stray field and ultrafast Néel vector dynamics at terahertz scale. The unique entanglement between the two magnetic sublattices and their alternating crystal environment in altermagnets provides unprecedented opportunities for designing their magnetic space symmetry. Many experimental efforts have been concentrated on tunning the Néel vector[7,13–17], in order to modulate the altermagnetic symmetry and the corresponding TRS-breaking responses such as anomalous Hall effect (AHE)[13–17] and spin-splitting texture[7], etc. However, the manipulation of altermagnetic symmetry through the crystal aspect has been neglected. Actually, it is quite challenging to realize the crystal design of altermagnetism, which means only changing the crystal symmetry of altermagnets while maintaining the Néel vector easy-axis. The key of a feasible scheme lies in modulating the crystallographic plane with a fragile high symmetry in the film setup, and the magnetic anisotropy of the Néel vector should be easy-axis instead of easy-plane, with the easy-axis being parallel to the high-symmetric crystallographic axis. In this case, different crystal distortions can efficiently induce different types of symmetry breaking.

As we know, magnetic exchange-coupling interactions reflect the essential characteristics of a magnetic phase. Therefore, in order to investigate the emergent physical properties induced by the crystal design of altermagnetism, it is indispensable to make theoretical analyses about the crystal-distortion-induced reconstruction of magnetic exchange-coupling interactions and their mutual relationship at first. In altermagnets, the intrinsic magnetic exchange-coupling interactions include antiferromagnetic exchange-coupling interaction and Dzyaloshinskii-Moriya (DM) interaction[34,35]. The former is a symmetric magnetic interaction described by a scalar $J_{AF}$. It is fundamental to the generation of antiferromagnetic exchange-coupling torque[18–20], which is an exotic innate spin torque distinct from the spin-orbit torque (SOT)[36–38]. Whereas DM interaction is a chiral exchange-coupling interaction described by a vector ***D***, originating from the combination of symmetry breaking and spin-orbit coupling.[4,24] Since different symmetry breaking can be realized by different crystal distortion, there is a locking between the crystal design of altermagnetic symmetry and the emergent DM vectors. The concept of exchange-coupling torques can be broadened to include both antiferromagnetic exchange-coupling torque and DM torque. They not only serve as powerful tools to provide us insight into the design of the relationship between these two types of exchange-coupling interactions in altermagnets, but also are fundamental to the electrical manipulation of altermagnetic order, which will inspire future works in extensive fields.

In this work, we propose CrSb as an appropriate material for realizing the crystal design of altermagnetism. CrSb is an altermagnetic metal with A-type antiferromagnetic spin configuration and a high altermagnetic transition temperature (710 K). The spin-splitting is as large as 1200 meV and beneficial for generating giant tunneling magnetoresistance.[10,22] We firstly catch a quick glimpse of the magnetic space symmetry of CrSb in the bulk state (Fig. 1a). Since the magnetic easy-axis for Néel vector points along [0001] crystallographic orientation, bulk CrSb belongs to magnetic space group (MSG) P6_3'/m'm'c. There is a highly symmetric crystallographic plane (0001), with itself as glide magnetic mirror plane (denoted as $M \cdot t_{1/2} \cdot T$), and three glide mirror planes ($M \cdot t_{1/2}$) as well as three magnetic mirror planes ($M \cdot T$) orthogonal to it. $\{11\bar{2}0\}$ are the magnetic mirror planes and $\{1\bar{1}00\}$ are the glide mirror planes, thus each magnetic mirror plane here has a glide mirror plane orthogonal to it. A magnetic mirror plane reverses the direction of a pseudovector orthogonal to it, while a glide mirror plane reverses the direction of a pseudovector parallel to it. As we know, both anomalous Hall (AH) vector and net magnetization are pseudovectors. Therefore, this kind of magnetic space symmetry prohibits anomalous Hall vector and net magnetization along any direction, leading to perfectly collinear Néel vector with zero DMI. To study the symmetry of SOT switching, we make a mirror reflection operation to the Néel vector, with the mirror plane parallel to the easy axis of Néel vector and orthogonal to current-induced spin polarization *p* (Fig. 1b). The time-reversal counterpart of the original Néel vector in region I is identical to its mirror reflection counterpart displayed in region II, prohibiting energy preference, and hence only continuous rotation rather than deterministic switching would occur.

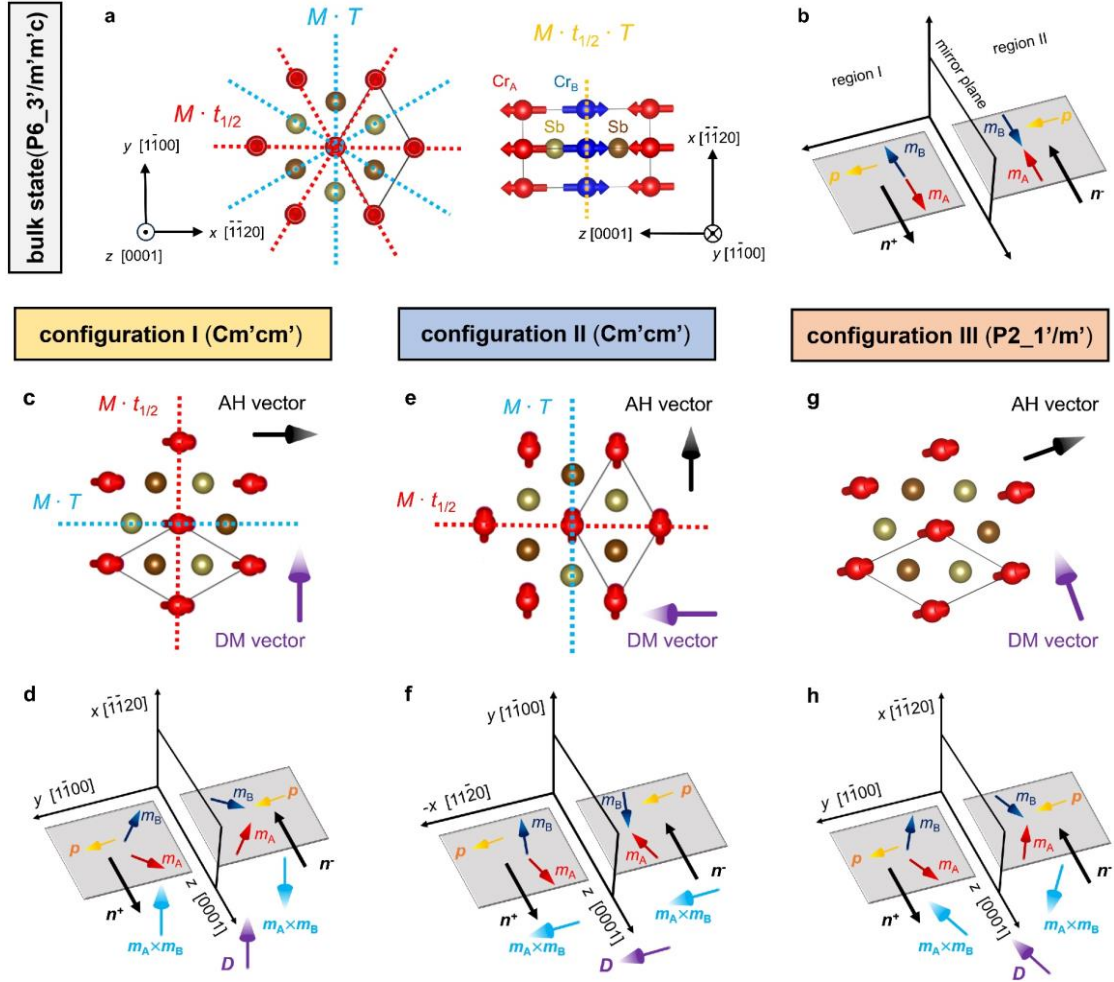

**Fig. 1 | Magnetic space symmetry and symmetry analyses of spontaneous AH vector and SOT switching modes for CrSb in the bulk state and thin film states. a, c, e, g**, Magnetic space symmetry of bulk CrSb (**a**), and thin films belonging to configuration I (**c**), II (**e**) and III (**g**). $M$ represents the mirror reflection operation. $t_{1/2}$ denotes the half-a-unit-cell translational operation along CrSb[0001]. $T$ is the time-reversal operation. Hereafter, CrSb[0001], [1$\bar{1}$00] and [$\bar{1}\bar{1}$20] is defined as $+z$, $+y$ and $+x$ direction, respectively. **b, d, f, h**, Symmetry analyses of SOT switching modes in bulk CrSb (**b**), and thin films belonging to configuration I (**d**), II (**f**) and III (**h**). $D$ and $p$ represent DM vector and current-induced spin polarization, respectively. The two magnetic sublattices are denoted as $m_A$ and $m_B$. Néel vector $n = (m_A - m_B)/2$. The 0 ° and 180 ° states of altermagnetic Néel vector $n$ are denoted as $n^+$ and $n^-$,

respectively.

The (0001) crystallographic plane of CrSb is in a fragilely protected highly-symmetric state, which is sensitive to crystal distortion, providing a unique opportunity for the symmetry design. The DM vector arises from the symmetry breaking, with the orientation dependent on the preserved magnetic space symmetry of CrSb thin films. Therefore, we study the generation of DM vector as well as its influence on AHE and SOT symmetry in the following three configurations (Fig. 1c–h). In configuration I, only one pair of glide mirror and magnetic mirror plane is preserved, and the normal axis of the film surface is parallel to the glide mirror plane (Fig. 1c). This lowers the MSG to Cm'cm', producing Dzyaloshinskii-Moriya vector (denoted as $D$) parallel to $[11\bar{2}0]$ crystallographic axis, which is also the normal axis of the surface in this configuration, with the net magnetization and AH vector here parallel to $[1\bar{1}00]$. The emergent DM vector makes the Néel vector deviate from the perfectly collinear state, leading to the emergence of a non-zero vector, $m_A \times m_B$ (Fig. 1d). In region I, $m_A \times m_B$ points to $[\bar{1}\bar{1}20]$, while its direction inverts after mirror reflection. The DMI energy of $n^+$ and $n^-$ states in this configuration are not equivalent ($E_{DMI} = -D \cdot (m_A \times m_B)$). The asymmetric DMI energy makes it possible for deterministic 180° switching without assistance of external field. Unfortunately, the AH vector parallel to the film surface cannot be observed experimentally, prohibiting the electrical readout of Néel vector switching. In configuration II, the preserved magnetic space symmetry is the same as that in configuration I, but the normal axis of the film surface is parallel to the magnetic mirror plane, yielding the AH vector orthogonal to the film surface and observable (Fig. 1e). In stark contrast to configuration I, $m_A \times m_B$ does not reverse direction under the mirror reflection, therefore the ***n***$^+$ and ***n***$^-$ states are equivalent in terms of energy in this configuration (Fig. 1f). Therefore, in order to realize 180° deterministic switching, an external magnetic field is required to break the magnetic mirror symmetry. In configuration III, there are no pairs of magnetic mirrors and glide mirrors, further lowering the MSG to P2_1'/m' (Fig. 1g). Configuration III can be viewed as the mixing of configuration I and II, and may realize both observable AHE and field-free switching (Fig. 1h).

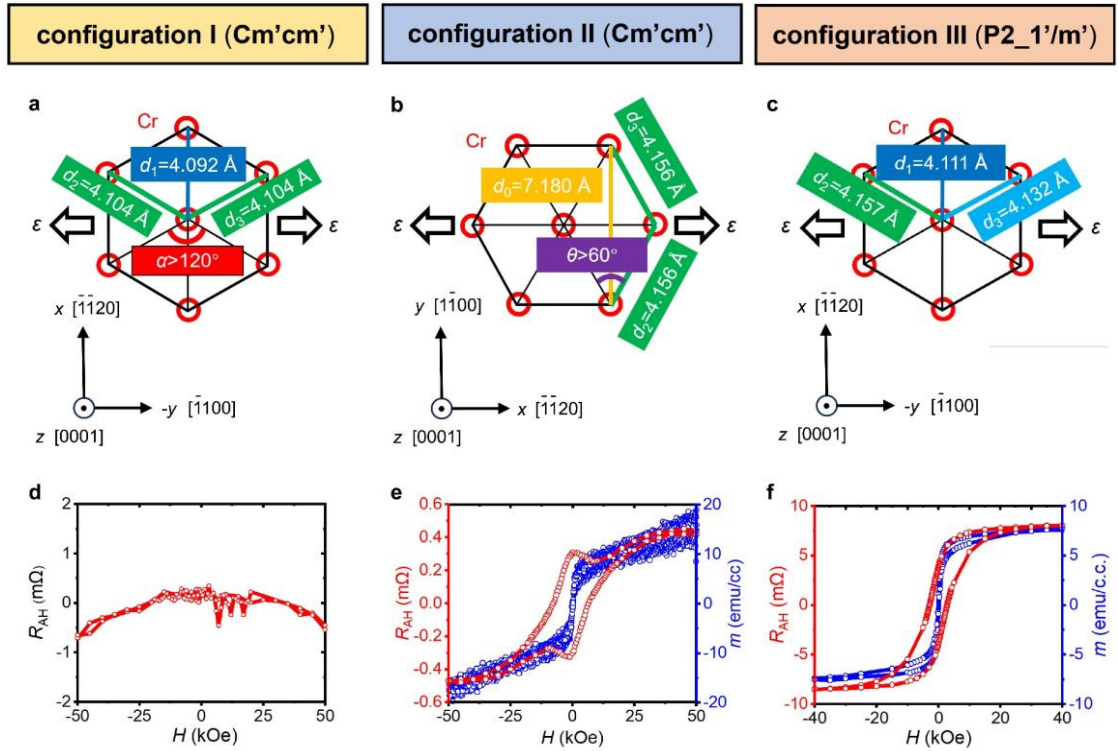

**Fig. 2 | Crystallographic parameters, room temperature AHE and out-of-plane magnetization measurement for strained CrSb films belonging to different configurations.** $d_1$, $d_2$ and $d_3$ are the length of crystallographic axis [11$\bar{2}$0], [$\bar{1}$2$\bar{1}$0] and [2$\bar{1}\bar{1}$0], respectively. $d_0$ denotes the length of the crystallographic axis [1$\bar{1}$00]. **a–c**, Crystallographic parameters of CrSb films in configuration I (**a**), II (**b**) and III (**c**). **d**, Anomalous Hall resistance ($R_{AH}$) of CrSb film belonging to configuration I. **e, f**, AHE and out-of-plane magnetization of CrSb films belonging to configuration II (**d**) and III (**f**).

The three configurations listed above can be realized experimentally via a combination of the uniaxial strain supplied by substrates and the growth control. The magnetic space symmetry of CrSb films is determined by the precise measurement of crystallographic parameters through performing conventional and off-axis x-ray diffraction (XRD) characterization. The crystallographic parameters of the three configurations are shown in Fig. 2a–c. Experimental setup and XRD data are displayed in Supplementary Note 1. For configuration I, we deposited 2.5 nm-thick W buffer on $Al_2O_3(11\bar{2}0)$ substrate for the growth of 30 nm-thick $CrSb(11\bar{2}0)$ film at 300 ℃, followed by 700 ℃ post-annealing. Figure 2a illustrates that the length of crystallographic axis $d_2$ and $d_3$ are larger than $d_1$, indicating the angle between $[2\bar{1}\bar{1}0]$ and $[\bar{1}2\bar{1}0]$, $\alpha$, is larger than 120 ° and there is tensile stress parallel to $[1\bar{1}00]$. Such tensile strain breaks the glide mirrors and magnetic mirrors tilted (not orthogonal, not parallel) to the film surface, as illustrated in Fig. 3c. But the relationship that $d_2 = d_3$ reflects the reservation of glide mirror orthogonal to and magnetic mirror parallel to the film surface, prohibiting the observable AH vector. Thus there is no AH hysteresis loop in Fig. 2d.

For configuration II, $CrSb(1\bar{1}00)$ film was grown directly on $LaAlO_3(110)$ substrate at 300 ℃, accompanied by 500 ℃ post-annealing and the growth of 5 nm Pt cap. The relationship of $d_2 = d_3$ and $d_0 < \sqrt{3}d_2$ makes the angle between $[2\bar{1}\bar{1}0]$ and $[1\bar{2}10]$, $\theta$, larger than 60 °, suggesting the existence of tensile stress parallel to $[11\bar{2}0]$, as marked in Fig. 2b. The tensile strain breaks the glide mirrors and magnetic mirrors tilted to the film surface, leaving only a magnetic mirror orthogonal to and a glide mirror parallel to the surface, as displayed in Fig. 1e. As expected, spontaneous AHE can be observed at room temperature, with a large coercive field of $H_c = 7$ kOe (Fig.

2e). Remarkably, this coercive field is greatly larger than that of the out-of-plane magnetization hysteresis loop (~50 Oe). The tiny net magnetization (~20 emu/c.c.) most likely arises from the unavoidable defects during the film growth. Both the mismatch of the coercivity and the tiny net magnetization indicate that AHE does not originate from ferromagnetism[13]. The vanishingly small intrinsic net magnetization means that CrSb keeps altermagnetic nature even though its magnetic space symmetry is changed. Interestingly, we show in Fig. 2c that $d_2$ is unequal to $d_3$ in CrSb($11\bar{2}0$) film when the growth condition is identical to configuration I but only with a lower growth temperature of 240 ℃ and annealing temperature of 580 ℃. In this scenario, sizable AHE emerges in the W/CrSb($11\bar{2}0$) film at room temperature, exhibiting the coercive field $H_c$ of ~2.8 kOe much larger than its magnetization counterpart (~50 Oe), as displayed in Fig. 2f. It is worthy mentioning here that the Néel vector rotates around the out-of-plane magnetic field during the AHE measurements (See Supplementary Note 2). In addition, AHE in both configuration II and III sustain up to 400 K, the upper limit of testing instrument (Supplementary Note 3), confirming the robustness of AHE in CrSb.

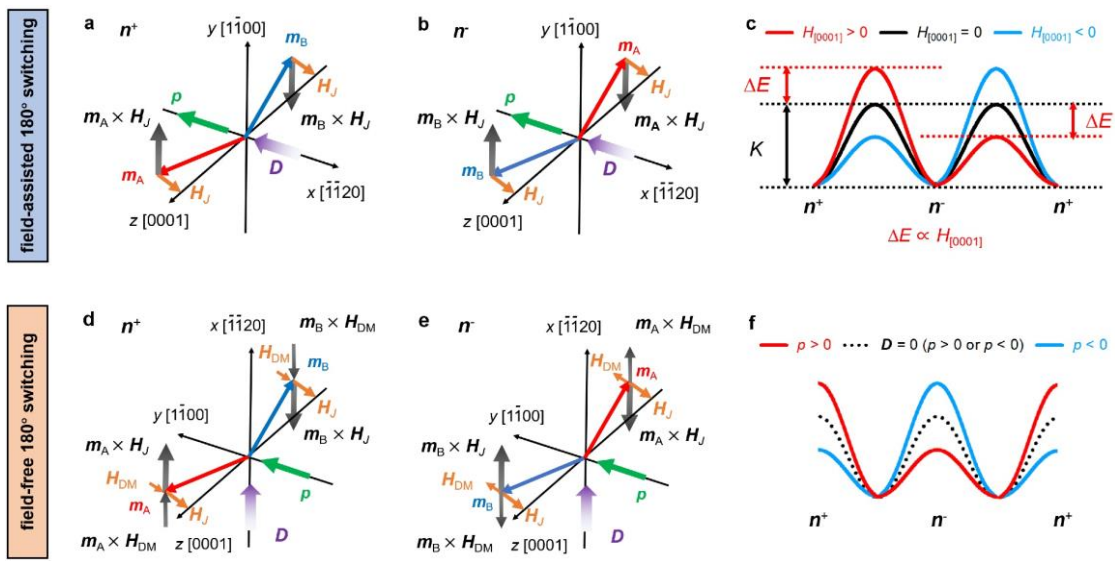

**Fig. 3 | Mechanisms of field-assisted (a–c) and field-free (d–f) 180 ° deterministic

**switching. a**, **b**, Schematic of effective fields and spin torques induced by exchange coupling, experienced by the two magnetic sublattices in the $n^+$ state (**a**) and $n^-$ state (**b**). **c**, Energy barrier landscape of Néel vector rotation at $H_{[0001]} = 0$ (black line), $H_{[0001]} < 0$ (blue line) and $H_{[0001]} > 0$ (red line). $K$ parameters the magnetic anisotropy energy, $\Delta E$ marks the energy barrier variation induced by Zeeman interaction. **d, e**, Schematic of effective fields and spin torques induced by exchange coupling, experienced by the two magnetic sublattices in the $n^+$ state (**d**) and $n^-$ state (**e**). **f**, Exchange-coupling torques landscape of Néel vector rotation. Black dotted line represents the exchange-coupling torques at the hypothetical zero $D$ situation. Red and blue lines denote the exchange-coupling torques under the positive (red line) and negative (blue line) spin polarization.

We now address the emergent exchange-coupling torques and electrical manipulation mechanisms in altermagnets. When the current-induced spin polarization $p$ is orthogonal to the Néel vector, it produces damping-like torques $m_{A/B} \times (p \times m_{A/B})$, causing the deviation of two magnetic sublattices from the equilibrium state and resultant exchange-coupling torques. The exchange coupling between the two magnetic sublattices of altermagnetic Néel vector can be described by the Hamiltonian H = $-J_{AF}\, m_A \cdot m_B - D \cdot (m_A \times m_B)$, where the first term is antiferromagnetic exchange coupling interaction and the second term stands for Dzyaloshinskii-Moriya interaction. These two kinds of interaction tends to produce the effective exchange interaction field as $H_{J, A/B} = J_{AF}\, m_{B/A}$ as well as $H_{DM, A} = -D \times m_B$ and $H_{DM, B} = +D \times m_A$ (See Methods for the derivation). The resultant field-like exchange-coupling torques $m \times (H_{J, A/B} + H_{DM, A/B})$ serve as the driving force of Néel vector rotation. Obviously, only the effective exchange interaction field components parallel (or antiparallel) to $p$ can drive the rotation, which are depicted in Fig. 3.

Now we compare the mechanisms responsible for the field-assisted and field-free switching in altermagnets. In the field-assisted mode, because of $D \parallel p$, the effective DMI field $H_{DM, A/B}$ is perpendicular to $p$. Only $H_{J, A/B}$ contributes to the Néel vector rotation. In this scenario, the driving force is symmetric irrespective of whether the Néel vector is in $n^+$ or $n^-$ state (Fig. 3a and b). Thus an asymmetric reconstruction of energy barrier is indispensable for the deterministic switching, which can be introduced by assistant magnetic field. The energy barrier variation is proportional to $H_{[0001]}$ (Fig. 3c), consistent to the reduced threshold current density $J_c$ with increasing assistant field in Fig. 4c. When $p$ and the assistant field are along $-x$ and $+z$ directions, respectively, the energy barrier for the rotation from $n^-$ to $n^+$ state is smaller than that from $n^+$ to $n^-$ state, leading to the deterministic switching from $n^-$ to $n^+$ state. Reversing the direction of either $p$ or $H_{[0001]}$ would invert the switching chirality. Differently, when $p \parallel n$, $H_{J, A/B}$ cannot generate Néel vector rotation.

For the field-free mode, as illustrated in Fig. 3d and e, $H_{J, A/B}$ and $H_{DM, A/B}$ are parallel or anti-parallel, thereby both contributing to the driving force for the Néel vector rotation, expressed as $m \times (H_{J, A/B} + H_{DM, A/B})$. At the $n^+$ state, $H_{J, A/B}$ and $H_{DM, A/B}$ point to the same direction (Fig. 3d), thereby boosting the rotation from $n^+$ to $n^-$ state. After being driven into the $n^-$ state, $H_{J, A/B}$ and $H_{DM, A/B}$ point towards the opposite direction (Fig. 3e), thereby hampering the rotation from $n^-$ to $n^+$ state. Figure 3f presents that the deterministic 180° field-free switching of Néel vector is actually ascribed to the asymmetric driving force, arising from the nonequivalent DMI energy of the $n^+$ and $n^-$ states (Fig. 1d and h). When reversing the direction of $p$, $H_{DM, A/B}$ plays an obstructive role in the rotation from $n^+$ to $n^-$ state, yet it helps the rotation from $n^-$ to $n^+$ state (Fig. 3f and Supplementary Note 4). Fig. 3f summarizes that when $D$ = 0, reversing the direction of $p$ cannot change the amplitude of exchange-coupling torques. However, once there is a non-zero $D$ orthogonal to $p$, the exchange-coupling torques are dependent on both the Néel vector state (for instance, $n^+$ or $n^-$) and the sign of $p$. Exchange-coupling torques are an intrinsic property determined by the magnetic space symmetry of altermagnetic CrSb. Clearly, the assistant field has negligible influence on the exchange-coupling torques. In addition, the high efficiencies in the two modes reflect the advantage of exploiting exchange-coupling torques in Néel vector switching in altermagnets. The asymmetric reconstruction of energy barrier (Fig. 3c) and asymmetric driving force (Fig. 3f), responsible for the field-assisted and field-free switching modes, respectively, can be unified in the framework of Landau-Lifshitz-Gilbert equation (See Methods for the derivation). The atomic spin dynamics simulation (Supplementary Note 5 and 6) also supports the theoretical analyses above.

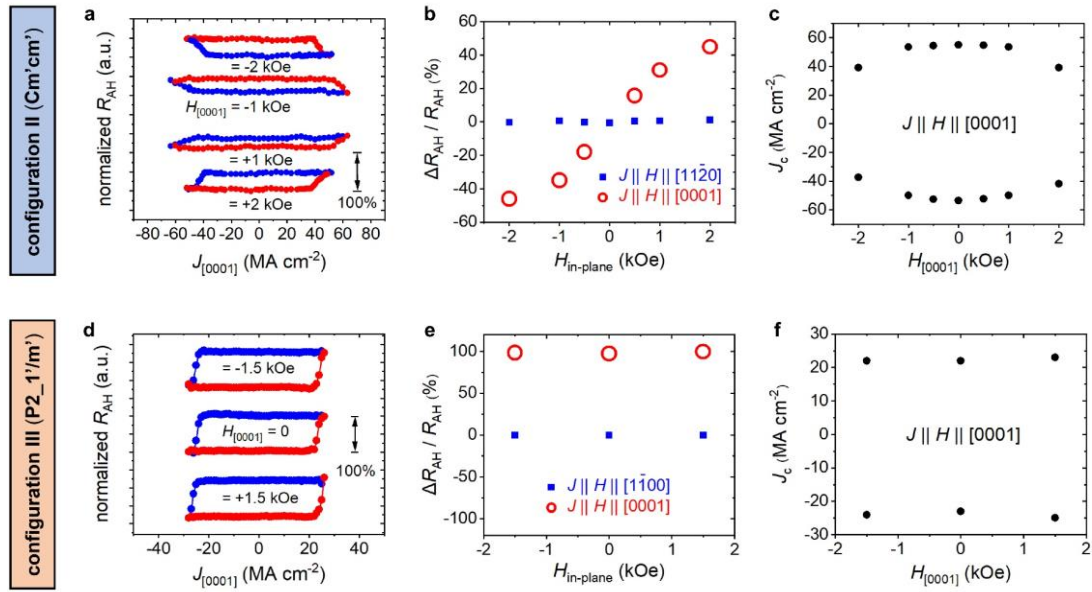

**Fig. 4 | Observation of exchange-coupling torques-induced 180° deterministic switching in CrSb(1$\bar{1}$00)/Pt (a–c) and W/CrSb(11$\bar{2}$0) (d–f). a, d,** Normalized $R_{AH}$-$J_{[0001]}$ curves under different assistant field $H$ applied along [0001] direction

($H_{[0001]}$). **b, e**, Dependence of switching ratio $\Delta R_{AH}/R_{AH}$ on the in-plane assistant magnetic field ($H_{\text{in-plane}}$) parallel to the electric current. **c, f**, Dependence of threshold current density $J_c$ on assistant field $H_{[0001]}$ with electric current flowing parallel to [0001].

We then turn towards electrical manipulation experiments for configuration II and III, with the readout scheme of AHE. We firstly get $R_{AH}$-$I$ curve, then $R_{AH}$ is normalized by the zero-field anomalous Hall resistance, and current density in the heavy metallic layer $J$ is obtained from $I$. Corresponding normalized $R_{AH}$-$J$ curves are displayed in Fig. 4. We present in Fig. 4a normalized $R_{AH}$-$J$ curves of CrSb(1$\bar{1}$00)/Pt film (configuration II) for $J \parallel$ [0001] and assistant field $H_{[0001]} = \pm 1$ and $\pm 2$ kOe. Clear hysteresis is observed with the switching polarity dependent on the sign of the assistant field. A summary of the assistant field dependent switching ratio $\Delta R_{AH}/R_{AH}$ in Fig. 4b displays the enhancement of $\Delta R_{AH}/R_{AH}$ from 18% to 45% as $H_{[0001]}$ increases from 0.5 to 2 kOe. This process is associated with the drop of threshold current density $J_c$, from ~54 to ~39 MA cm$^{-2}$, as illustrated in Fig. 4c. Normalized $R_{AH}$-$J$ curve at $H_{[0001]} = 0$ exhibits a resistance transition to an intermediate level above a certain threshold current density, regardless of the polarity of initialization and current (See Supplementary Note 7). The threshold current density of ~56 MA cm$^{-2}$, also denoted as $J_c$, is included in Fig. 4c. There is no deterministic switching for $H_{[0001]} = 0$, demonstrating that the assistant field is indispensable for configuration II. Note that $J_c$ at zero assistant field is larger than that for deterministic switching. When the current is along [11$\bar{2}$0], however, the transition does not occur, even though the pulse amplitude is up to 80 MA cm$^{-2}$. Even the assistant field is applied, the current flowing along [11$\bar{2}$0] cannot generate deterministic switching (Fig. 4b). The current-induced variation of $R_{AH}$, under both zero and non-zero assistant field, shows a strong crystallographic orientation dependence, supporting the electrical manipulation mechanism of altermagnets, where only the current-induced spin polarization that is orthogonal to the Néel vector works.

Identical measurements were performed for configuration III. The most eminent result in Fig. 4d is the deterministic full switching (~100%) occurs at zero assistant field for $J \parallel [0001]$. Also visible is that the switching polarity does not reverse as the sign of the assistant field changes (Fig. 4d and e). Figure 4f presents that the threshold current density $J_c$ is insensitive to the assistant field. These unique features support the novel mechanism responsible for the electrical manipulation in configuration III, where the DM interaction is determined by MSG and the total exchange-coupling torques show immunity to the assistant magnetic field. We then analyze $H_c/J_c$, which has been used as a figure of merit to evaluate the efficiency to manipulate magnetic orders. The switching in both configurations is quite efficient with $\mu_0 H_c/J_c = 0.0197$ T cm$^2$ MA$^{-1}$ at $H_{[0001]} = 2$ kOe for configuration II and $\mu_0 H_c/J_c = 0.0133$ T cm$^2$ MA$^{-1}$ for configuration III. These values are even one order of magnitude larger than that in ferromagnets (~0.001 T cm$^2$ MA$^{-1}$), and is comparable to that in ferrimagnets (0.001–0.02 T cm$^2$ MA$^{-1}$)[18,19] and noncollinear antiferromagnets (0.01–0.03 T cm$^2$ MA$^{-1}$)[39–41].

In conclusion, we realize the crystal design of altermagnetism and make a close attachment to the magnetic space symmetry in altermagnets to the emergent DM vectors, bringing about room-temperature spontaneous AHE in altermagnets or even collinear antiferromagnets, which is a long-sought goal in spintronics. The concept of exchange-coupling torques is broadened. We design the relationship between antiferromagnetic exchange-coupling torque and DM torque, and thereby design the electrical manipulation modes in altermagnets. Particularly, the 100-percent field-free switching of Néel vector is observed. This is critical for the practical use of high-density memories, having no need of accommodating a magnetic field. The crystal design of altermagnetism can be extended to a wide range of materials, with multiple magnetic mirrors and glide mirrors emerging in a hexagon-like lattice, and the Néel vector easy-axis orients parallel to the high-symmetric axis (see Supplementary Note 8). The DM interaction plays an essential role in magnonics, including magnon topological phases[23], fertile magnonic modes[42–48], as well as the magnonic transport[42] with the features of non-dissipation[49,50], non-reciprocity[50,51] and non-volatility[42,50]. Therefore, the crystal design of DM vectors in altermagnets would bring rich designability to magnonics.

**Methods**

**Sample preparation and characterization.** The CrSb films were prepared by magnetron sputtering with a rate of 1.6 Å s$^{-1}$ under a base pressure below $5 \times 10^{-8}$ Torr. For growing CrSb(11$\bar{2}$0), W buffer layer was deposited firstly to help the growth of CrSb film. Al$_2$O$_3$(11$\bar{2}$0) substrate was pre-annealed at 800 °C for 30 minutes, then W buffer layer with thickness as 2.5 nm was grown at 680 °C with a rate of 0.16 Å s$^{-1}$, followed by annealing at 800 °C for 30 minutes. Next, the 30 nm CrSb film was deposited by cosputtering with Cr and Sb targets, followed by post-annealing for 30 minutes. For the CrSb(11$\bar{2}$0) film belonging to configuration I, the depositing temperature of CrSb is 300 °C and the post-annealing is at 700 °C, while for the CrSb(11$\bar{2}$0) film belonging to configuration III, the depositing temperature of CrSb is 240 °C and the the post-annealing is at 580 °C. For the growth of CrSb(1$\bar{1}$00), LaAlO$_3$(110) substrate was pre-annealed at 500 °C for 30 minutes, then the 30 nm CrSb film was deposited onto the substrate at 300 °C, followed by post-annealing at 500 °C for 30 minutes. After that, 5 nm Pt was deposited in-situ at room temperature, with a rate of 0.2 Å s$^{-1}$. Cross bars with width as 5 μm for SOT switching measurements were patterned using optical lithography combined with Ar ion milling.

**Characterizations.** The *θ–2θ* XRD characterizations at room temperature was carried out at Rigaku Smartlab. Experimental setup for the coventional and the off-axis XRD characterizations is shown in Supplementary Note 1. Magnetic hysteresis curves were collected in commercial superconducting quantum interference device (SQUID, Quantum Design) and vibrating sample magnetometer of commercial Physical Property Measurement System (PPMS–VSM, Quantum Design), where the diamagnetic contributions of substrates were subtracted.

**Transport measurements.** Anomalous Hall resistance of CrSb films were measured in commercial Physical Property Measurement System (PPMS). The contribution from ordinary Hall resistance that is linear to magnetic field was subtracted. For SOT switching measurements, 1.5 ms writing pulses were added, followed by waiting for 10 s before collecting the Hall voltage by a nanovoltmeter.

**Derivation of spin torques experienced by the two magnetic sublattices in altermagnetic CrSb**. The Hamiltonian describing the altermagnetic Néel vector in CrSb is

$$H_{total} = -J_{AF}(\bm{m}_A \cdot \bm{m}_B) - \bm{D} \cdot (\bm{m}_A \times \bm{m}_B) - K(\bm{k}_A \cdot \bm{m}_A)^2 - K(\bm{k}_B \cdot \bm{m}_B)^2 - \mu_0(\bm{m}_A + \bm{m}_B) \cdot \bm{H}_{ext} \quad (1)$$

where $\bm{m}_A$ and $\bm{m}_B$ are the two magnetic sublattices, the first and second terms of the Hamiltonian are two different types of exchange coupling interaction, the third and fourth term are the magnetic anisotropy energy term, and the last term is the Zeeman interaction term. The two types of exchange coupling interaction are the symmetric antiferromagnetic exchange coupling interaction, and the antisymmetric one named as Dzyaloshinskii-Moriya interaction. $J_{AF} < 0$, $\bm{D}$ stands for DM vector, $K$ is the strength of single-ion anisotropy, and the unit vector $\bm{k}_A$ and $\bm{k}_B$ are the single-ion easy-axis parallel to CrSb[0001], $\mu_0$ is permeability of vacuum and $\bm{H}_{ext}$ represents external magnetic field. The Landau-Lifshitz-Gilbert equation[54] of the two magnetic sublattices in CrSb is expressed as

$$d\,m_{A/B}/dt = -m_{A/B} \times H_{A/B}^{eff} - \alpha\, m_{A/B} \times d\,m_{A/B}/dt + T_{A/B} \quad (2)$$

where $\alpha$ is the magnetic damping, $H_{A/B}^{eff}$ represents the effective magnetic fields experienced by sublattice $m_{A/B}$, and $T_{A/B}$ stands for spin-orbit torque exerting on sublattice $m_{A/B}$ induced by the current-induced polarization $p$ from heavy metal layer. For simplicity yet not losing universality, we only consider the damping-like spin-orbit torque, thus $T_{A/B}$ is proportional to $m_{A/B} \times (p \times m_{A/B})$. $H_{A/B}^{eff}$ are obtained from the functional derivative of the Hamiltonian,

$$H_A^{eff} = -\delta H_{total}/\delta m_A = H_{J,A} + H_{DM,A} + H_{K,A} + H_{Zeeman,A} \quad (3)$$
$$= J_{AF}\, m_B - D \times m_B + K(k_A \cdot m_A)\,k_A - \mu_0 H_{ext}$$

$$H_B^{eff} = -\delta H_{total}/\delta m_B = H_{J,B} + H_{DM,B} + H_{K,B} + H_{Zeeman,B} \quad (4)$$
$$= J_{AF}\, m_A + D \times m_A + K(k_B \cdot m_B)\,k_B - \mu_0 H_{ext}$$

Herein, we divide the effective magnetic fields $H_{A/B}^{eff}$ into two parts. The combination of the first term $H_{J,A/B}$ and the second term $H_{DM,A/B}$ contributes to the driving force for the altermagnetic Néel vector rotation, while the combination of the third term $H_{K,A/B}$ and the fourth term $H_{Zeeman,A/B}$ accounts for the energy barrier for the SOT-induced Néel vector rotation. Only the effective exchange coupling interaction fields that are parallel or antiparallel to the current-induced spin polarization $p$ could contribute to driving the Néel vector rotation in altermagnets. Thus, we only draw the components of effective fields that are parallel or antiparallel to $p$ in Fig. 3.

ferromagnetic materials. *IEEE Trans. Magn.* **40**, 3443–3449 (2004).

**Acknowledgments**

This work is supported by National Natural Science Foundation of China (Grant No. 52225106, 12241404, T2394470, and 12022416), the National Key Research and Development Program of China (Grant No. 2022YFA1402603, 2021YFB3601301, and 2021YFA1401500), and Hong Kong Research Grants Council (16303821, 16306722 and 16304523).